\documentclass[aps, reprint]{revtex4-1}

\pdfoutput=1
\usepackage{amsmath}    
\usepackage{graphicx}   
\usepackage{verbatim}   
\usepackage{color}      
\usepackage{subfigure}  
\usepackage{upgreek}

\usepackage{fancyhdr}
\begin{document}

\title{Genetic algorithm-based control of birefringent filtering\\for self-tuning, self-pulsing fiber lasers}
\author{R. I. Woodward and E. J. R. Kelleher}
\affiliation{Femtosecond Optics Group, Photon Science Section, Department of Physics, Blackett Laboratory, Imperial College London, London, UK}

\begin{abstract}
Polarization-based filtering in fiber lasers is well-known to enable spectral tunability and a wide range of dynamical operating states.
This effect is rarely exploited in practical systems, however, because optimization of cavity parameters is non-trivial and evolves due to environmental sensitivity.
Here, we report a genetic algorithm-based approach, utilizing electronic control of the cavity transfer function, to autonomously achieve broad wavelength tuning and the generation of Q-switched pulses with variable repetition rate and duration.
The practicalities and limitations of simultaneous spectral and temporal self-tuning from a simple fiber laser are discussed, paving the way to on-demand laser properties through algorithmic control and machine learning schemes.
\end{abstract}

\maketitle
\thispagestyle{fancy}

Fiber lasers are an important technology that continues to enable new applications as device performance, flexibility and reliability improve.
Alongside research pushing the frontiers of laser specifications in the laboratory, there is a need to develop tunable, turn-key systems for non-expert users, allowing precise control of temporal and spectral properties of the source.

Spectral tunability can be achieved by introducing a bulk interference or birefringent filter, or diffraction grating into the cavity.
Free-space components, however, eliminate the alignment-free benefits of an all-fiber system.
A solution is to use an \textit{artificial} birefringent filter, formed from the combined effect of dispersive linear polarization rotation due to fiber birefringence and polarization-dependent loss~\cite{Humphrey1993}.
This is analogous to a Lyot filter, exhibiting a comb-like transmission function.

Duration-tunable pulses can be generated by Q-switching (QS) and/or mode-locking using an electrically-driven modulator.
A conceptually simpler approach, however, is passive pulse generation exploiting nonlinearity in fiber. 
Nonlinear polarization rotation (NPR), for example, manifests itself through an intensity-dependent change in the polarization state of propagating light, yielding a power-dependent transmission when combined with a polarizer---i.e. forming an \emph{effective} saturable absorber (SA) through nonlinear birefringent filtering.
The transfer function can be modified through polarization control, varying the modulation depth and saturation intensity, and enabling a range of pulsed behaviors.
Despite progress in the development of new \textit{real} SA materials~\cite{Woodward2015_as_2d}, \textit{artificial} SAs remain a robust approach to pulse generation, with a quasi-instantaneous response and without requiring advanced material fabrication.

Exploitation of linear and nonlinear polarization-based filtering in all-fiber cavities has enabled wide tunability over a range of output properties, including: $>$75~nm mode-locked tuning range~\cite{Meng2015}, multiwavelength operation with up to 25 distinct wavelengths simultaneously~\cite{Zhang2008}; and temporal states from CW to Q-switching and mode-locking.
More recent applications of birefringent filtering in all-normal dispersion lasers have stabilized pulse formation~\cite{Ozgoren2010}, and enabled tuning of both wavelength~\cite{Zhang2012_comb, Fedotov2012} and pulse duration~\cite{Fedotov2012} by exploiting the chirped pulse dynamics.

The use of polarization-based filtering techniques in practical systems has been limited, however, by two major problems.
Firstly, fiber birefringence can be modified by its environment  (i.e. thermally/mechanically induced random fluctuations), thus optimum polarization settings can drift, requiring regular readjustment.
Additionally, the coupled and nonlinear dependence of the output properties on cavity parameters that adjust temporal and spectral modulation result in a complex, time-consuming optimization procedure to achieve a desired output~\cite{Meng2015, Fedotov2012}.

To address this, automated parameter control approaches have been proposed, focused principally on achieving stable fixed-wavelength mode-locking~\cite{Radnatarov2013, Iegorov2016, Shen2012, Hellwig2010, Olivier2015}.
Algorithms that simply search for optimum states, however, struggle with the diversity of laser operating regimes that include multiple points of local maxima (in addition to the desired global maxima).
A promising and versatile solution to this problem is to employ machine learning~\cite{Fu2013, Brunton2014, Andral2015,Andral2016, Woodward_scirep_2016}, broadly defined as system development to perform given functions without explicit instruction.
Learning~\cite{Brunton2014} and genetic algorithms (GAs)~\cite{Andral2015,Andral2016, Woodward_scirep_2016} have recently been applied to fiber lasers to intelligently explore parameter space and locate global optima, e.g. stable single-pulse mode-locking~\cite{Woodward_scirep_2016}.

Here, we extend this approach by using a GA to automate self-tuning of a fiber laser to achieve user-specified temporal and spectral properties, by harnessing birefringent filtering.
Using automated polarization and pump power control, and designing appropriate fitness functions, our system is able to self-tune its wavelength over 55~nm and achieve self-Q-switching with target pulse properties in a 25~kHz repetition rate and 30~$\upmu$s duration range.
We explore the feasibility of simultaneous wavelength and repetition rate tunability and discuss the practicalities and limitations of self-tuning laser technology. 

\begin{figure}[bt]
	\centering
	\includegraphics[width=0.9\linewidth]{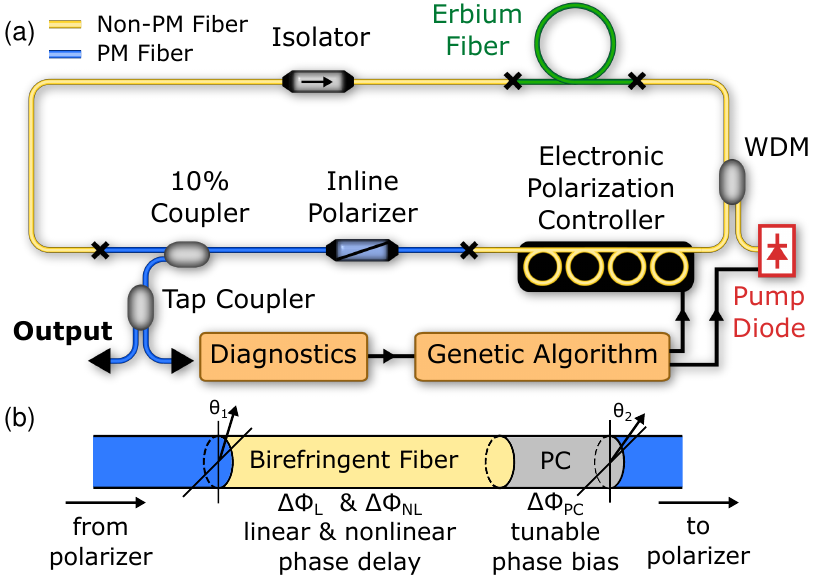}
	\caption{Self-tuning laser cavity: (a) schematic; (b) simplified illustration of phase delays arising from fiber birefringence.}
	\label{fig:cavity}
\end{figure}

A ring fiber laser design is used [Fig.~\ref{fig:cavity}(a)], including a 2.3~m length of erbium-ytterbium co-doped double-clad fiber, pumped at 965~nm, an electronic polarization controller (PC) consisting of four stepper-motor-driven quarter waveplates (QWPs) formed of fiber loops with stress-induced birefringence, and an in-line fiber polarizer.
The polarizer's fiber pigtails and subsequent 10\% output coupler are constructed from high-birefringence (i.e. polarization-maintaining, PM) fiber, ensuring a fixed output polarization from the laser.
All other fiber is low-birefringence (non-PM) Corning SMF28.
Finally, a polarization-independent in-line isolator is used, resulting in a total cavity length of 21~m.

To explain the operation of the artificial birefringent filter, we use a simplified analysis [Fig.~\ref{fig:cavity}(b)] considering the power transmission $T$ of the polarizer, which blocks light (with $>$26~dB extinction) polarized in the direction aligned with the fast-axis of its PM fiber pigtails.
Light in PM fiber after the polarizer remains linearly polarized in the fiber's slow axis.
Non-PM cavity fiber has an intrinsic birefringence orders of magnitude lower than PM fiber, although stresses from twisting and spooling non-PM fiber can significantly increase this.
The orthogonal polarization mode axes of the PM fibers form arbitrary angles $\theta_1$ and $\theta_2$ to the principal polarization axes of the non-PM fibers.
Light launched into non-PM fiber can excite both supported polarization modes, which couple and transfer power on propagation.
This causes a rotation of the polarization state through both a linear ($\Delta\phi_{L}= 2 \pi L \Delta n / \lambda$) and nonlinear ($\Delta\phi_{NL}= (2/3) \gamma L P \cos 2\theta_1$) phase shift between the two polarization components~\cite{Man2000}; where the fiber has length $L$, birefringence $\Delta n$, nonlinearity parameter $\gamma$, and the instantaneous optical power is $P$.
An additional phase shift term $\Delta\phi_{PC}$, variable by the user, arises from the PC, which permits complete traversal of the Poincaré sphere through the adjustable angles of four QWPs.
Light launched into the $\sim$0.2~m PM input fiber pigtail of the polarizer is not necessarily aligned to one of the principal axes, hence birefringence here also results in a phase delay, which for simplicity is assumed to be included in $\Delta\phi_{L}$ and $\Delta\phi_{NL}$ terms.
After a cavity round-trip, any light that is rotated to align with the polarizer's fast axis is attenuated, giving the power transmission function~\cite{Man2000}:
\begin{multline}
\label{eqn:trans}
	T = \cos^2\theta_1 \cos^2\theta_2 + \sin^2\theta_1 \sin^2\theta_2 \\ + 
	\frac{1}{2} \sin 2\theta_1 \sin 2\theta_2 \cos(\Delta \phi_{L} + \Delta \phi_{NL} + \Delta \phi_{PC}).
\end{multline}

\pagestyle{plain}

The final cosine term indicates that the birefringent filter transmission is periodic with wavelength due to $\Delta\phi_{L}$.
Spectral filtering is also introduced by weak wavelength-dependence of both $\gamma$ and $\Delta n$.
Importantly, the transmitted wavelengths can be tuned by adjusting the PC (i.e. $\Delta\phi_{PC}$).
In combination with the dynamic gain profile of the erbium fiber amplifier~\cite{Meng2015}, the birefringent filter defines the laser wavelength.  
To achieve pulsation, the phase bias $\Delta\phi_{PC}$ is set such that the linear transmission (i.e. for CW light) is low, while the filter permits a higher transmission for light of greater intensity, controlled by $\Delta\phi_{NL}$.
The modulation depth, non-saturable loss and saturation intensity of this artificial SA can thus be adjusted by the phase bias.

The identification of system parameters (here, the 4 PC waveplate angles and pump power) to achieve on-demand output properties is a non-trivial optimization.
Fortunately, GAs are ideally suited to  this task.
Briefly, GAs efficiently perform global multivariate optimization using principles from evolutionary biology to find parameters that maximize a quality score, as explained in detail for this context in Ref.~\cite{Woodward_scirep_2016}. 
The process starts with a population of random parameter sets, which are each trialled and scored according to a fitness function.
A new generation of parameter sets is then generated by `breeding' from the previous generation, where `parent selection' is probabilistic, related to the parent's score.
`Mutation' is applied by randomly varying parameters with a small probability, to prevent the algorithm converging to local optima. 
The new generation is then tested and the procedure repeats, identifying and maintaining the best parameters, while rejecting low-scoring sets.

We first demonstrate wavelength tuning, initially neglecting the temporal properties, to elucidate the GA evolution.
The fitness function, $F_\lambda = 1 - \frac{|\lambda - \lambda_0|}{0.5\Delta\lambda}$, is used to score the spectrum ($F_\lambda=1$ is optimal) measured on an optical spectrum analyzer (OSA), where $\lambda$ is the central wavelength (found using a peak detection routine on the recorded data; if no peak found, i.e. no lasing, $F_\lambda=0$), $\lambda_0$ is the target wavelength, and $\Delta\lambda=70$~nm is the estimated gain bandwidth.

\begin{figure}[tb]
	\centering
 	\includegraphics[width=0.95\linewidth]{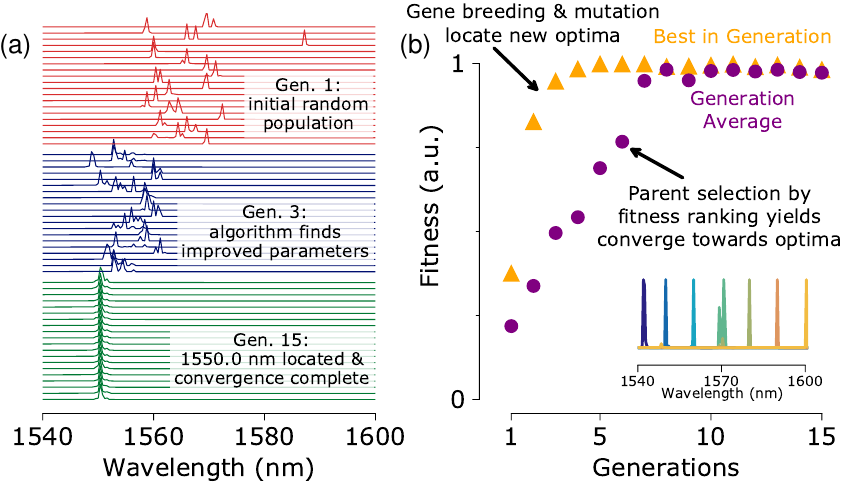}
	\caption{GA-based wavelength self-tuning: (a) visualization of spectra for each parameter set in generations 1, 3 and 15 (b) fitness evolution showing convergence towards optimum (inset: self-tuned spectra showing tuning range).}
	\label{fig:wl_tuning}
\end{figure}

The GA optimization procedure for a target wavelength of 1550.0~nm is visualized in Fig.~\ref{fig:wl_tuning}: (a) shows the 1st, 3rd and 15th generation results---the initial population of randomized parameters yields a random selection of laser wavelengths; subsequent generations are formed from crossover of the `best' parameters from the earlier generation, including a random mutation probability, resulting in the breeding out of `bad' parameters that yield laser wavelengths far from the target, and introducing new parameters with laser wavelengths closer to the target. 
After several generations, the algorithm converges to locate the optimum PC and pump power settings that yield the desired laser wavelength [with the evolution of the best and average generation score shown in Fig.~\ref{fig:wl_tuning}(b)]. 
We executed the algorithm with a range of target wavelengths: the tuning limits were found to be 1542.2 to 1600.4 nm.
Within this range, autonomous self-tuning of the birefringent filter was always successful, achieving on-demand wavelength selection within 0.1~nm of the target [Fig.~\ref{fig:wl_tuning}(b) inset].
The factor limiting this range is believed to be insertion losses in fixed fiber components (e.g. isolator/WDMs), which are specified for operation around 1560~nm.

The choice of GA parameters (generation size, crossover rate etc.~\cite{Melanie1996}) defines the accuracy and convergence rate of the system---i.e. time taken to locate the desired state.
Empirically, we achieved good performance using a generation size of 20, crossover rate of 90\%, and mutation rate of 20\% (which is linearly damped as the average score approaches 1).
For parent selection during breeding, we use a `rank selection' algorithm, in contrast to our previous GA work to find  mode-locking~\cite{Woodward_scirep_2016} using `roulette wheel selection'.
Roulette wheel selection assigns the selection probability based on a parameter set's score, favourable for identifying small mode-locked states within a wide parameter space of non-lasing, CW and QS regimes exhibiting high performance contrast.
Rank selection, however, chooses parameters based on their position in the fitness-sorted list of all parameter sets.
This enhances the contrast between similarly scoring parameter sets, and is thus better suited for implementing on-demand tunability over a continuous range.

\begin{figure}[tb]
	\centering
	\includegraphics[width=0.95\linewidth]{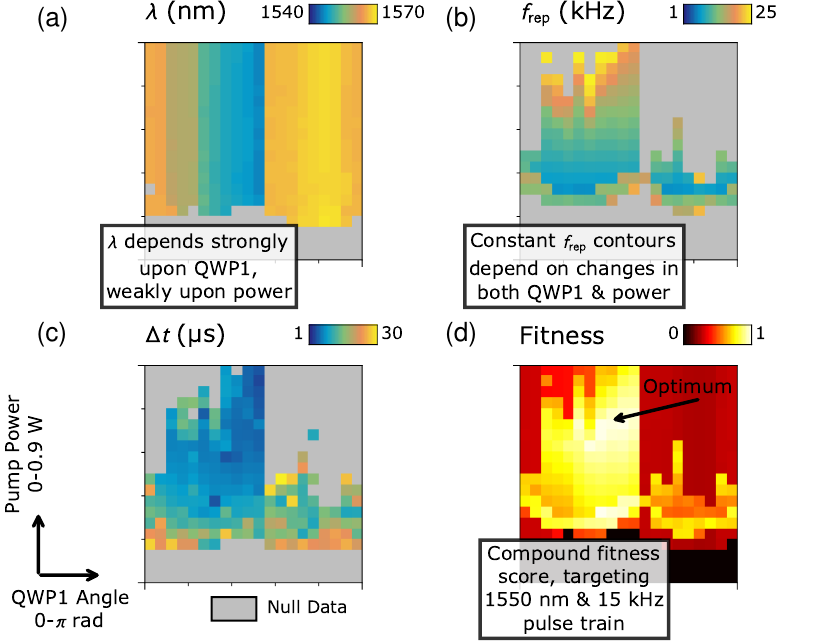}
	\caption{Maps of laser characteristics with respect to QWP1 angle (x-axis: 0 to $\pi$ rad) and pump power (y-axis: 0 to 0.9~W): (a) wavelength; (b) Q-switched repetition rate; (c) pulse duration; (d) score with targets: 1550~nm \& 15~kHz.}
	\label{fig:maps}
\end{figure}

We now consider temporal properties. 
To highlight the diversity of possible output states and the dependence on input parameters, we measure a two-dimensional slice of the five-dimensional parameter space: three QWPs are fixed, while the pump power is swept from 0 to 0.9~W and one QWP (hereafter, called QWP1) is swept through 180$^\circ$ [Fig. \ref{fig:maps}(a)-(c)]. 
At each point, the wavelength is measured, and when pulsing, the pulse duration and repetition rate are measured using a photodetector and oscilloscope.
The laser threshold is $\sim$$0.2$~W.
For many QWP1 settings above threshold stable QS is observed, with the pulse properties determined by the pump power and polarization phase bias.
Increasing the pump power during QS operation with fixed waveplates results in a linear increase in repetition rate [Fig. \ref{fig:maps}(b)] and decrease in pulse duration [Fig. \ref{fig:maps}(c)].
This is expected as a QS pulse is emitted once sufficient stored energy in the cavity is accumulated; thus, higher pump power enables increased repetition rates and shorter pulses.
Pulse properties are also affected by the QWP angle as this adjusts $\Delta \phi_{PC}$ in the birefringent filter, which varies the intensity-dependent-loss function.
To target autonomous control of a QS output with on-demand repetition rate (excluding wavelength tuning), we redefine the GA fitness function as $F_f = 1 - \frac{|f - f_0|}{f_0}$, with measured repetition frequency $f$ and target $f_0$, achieving self-tuning, self-pulsation over a 25~kHz repetition rate range.

Fig.~\ref{fig:maps} highlights that various combinations of laser properties can be accessed---e.g. a contour of constant repetition rate $\sim$15~kHz in Fig.~\ref{fig:maps}(b) corresponds to a region in Fig.~\ref{fig:maps}(a) where the wavelength varies over $\sim$10~nm. 
This shows that by adjusting just the power and QWP1, a 15~kHz pulse train could be achieved at any target wavelength in this range.
The tuning range here, however, by varying only two parameters is limited compared to the full performance range that can be achieved through varying all five parameters [e.g. 30~nm wavelength range in Fig.~\ref{fig:maps}(a) compared to 58~nm range in Fig.~\ref{fig:wl_tuning}].
An important question, therefore, is can we arbitrarily select wavelength, pulse duration and repetition rate within the full tuning range?

We explore the potential for this self-tunability using a GA with a compound fitness function that assigns a score according to both the wavelength and pulse repetition rate: $F_\mathrm{total}=0.5F_\lambda + 0.5F_f$.
The effect of this function in identifying a target operating regime (arbitrarily chosen to be $\lambda_0=1550$~nm wavelength, $f_0=15$~kHz repetition rate) is illustrated by applying it to the reduced 2D parameter space (of pump power and QWP1), resulting in the fitness map Fig.~\ref{fig:maps}(d), highlighting the optimum operation region for these properties.

\begin{figure}[tb]
	\centering
	\includegraphics[width=0.95\linewidth]{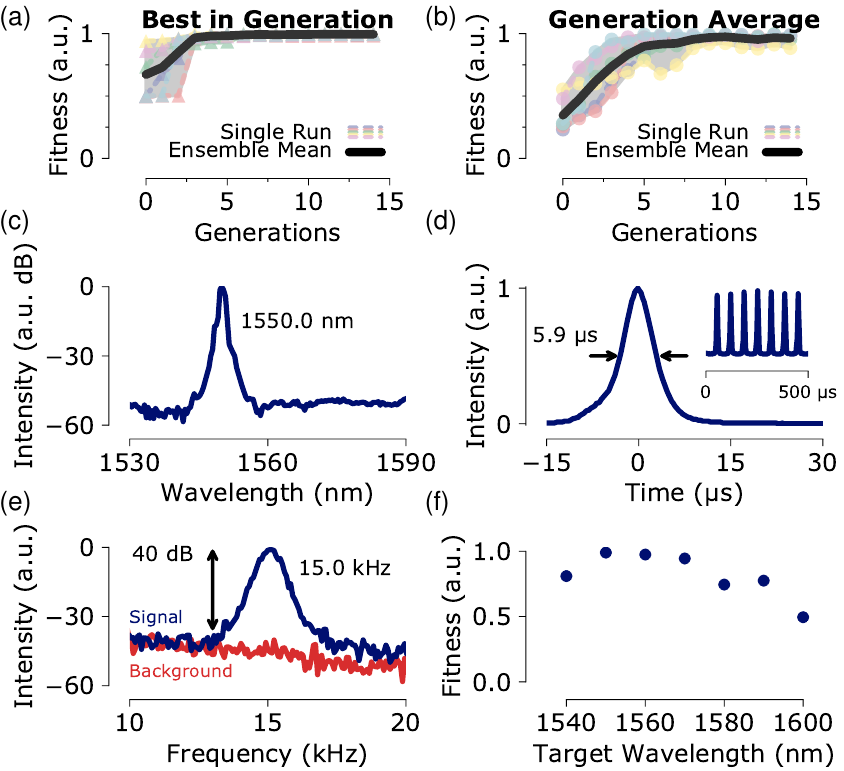}
	\caption{Self-tuning characteristics, with targets  $\lambda_0=1550.0$~nm and $f_0=15$~kHz: evolution of (a) 'best' and (b) 'average' fitness; performance with the self-optimized parameters, (c) optical spectrum, (d) pulse, (e) RF spectrum; (f) optimum achieved fitness with 15~kHz target repetition rate and variable target $\lambda$.}
	\label{fig:both_tuning}
\end{figure}

Self-tuning (from randomized initial conditions, including all five parameters) towards the target $\lambda_0=1550$~nm and $f_0=15$~kHz is demonstrated by running the compound fitness-function-based GA.
Within the first generation, a range of non-lasing, CW and QS states (with widely varying repetition rates) are observed, as expected from the broad parameter space (Fig.~\ref{fig:maps}), giving a low average score. 
The algorithm maintains the 'good' parameters and breeds/mutates them to identify improved performance---shown by the increase in the `Best in Generation', which leads to convergence to optimal performance over numerous generations [Figs.~\ref{fig:both_tuning}(a)-(b)]. 
The routine is re-executed, each time following a different evolution due to the probabilistic process and randomized initial conditions, but demonstrating reliability by always converging on the generation of a 1550~nm wavelength, 15~kHz pulse train---as characterized in Figs.~\ref{fig:both_tuning}(c)-(e).
The pulses exhibit a 5.9~$\upmu$s duration and the electrical spectrum shows a high peak-to-background contrast of 40~dB at the fundamental repetition frequency of the pulse train, indicating good stability. 
The 4.7~mW output is linearly polarized with a 19~dB extinction ratio, corresponding to 0.31~$\upmu$J pulse energy. 
We also note that when the laser is intentionally disturbed, causing the performance to change randomly, the GA quickly relocates new optimum parameters to restore the desired output properties (similar to Ref.~\cite{Woodward_scirep_2016}).

To explore the limits of this approach, we repeated the optimization process targeting 15~kHz pulse train generation over a range of wavelengths.
The optimal converged fitness values as a function of target wavelength are shown in Fig.~\ref{fig:both_tuning}(f).
Unity fitness is not achieved across the full range, highlighting that the target properties were not found in all cases. 
Consistently, high performing regimes are observed between 1550 and 1570~nm, with achieved optimal scores exceeding 0.96.
Outside of this region, however, the maximum score, despite numerous iterations of the self-tuning process, is lower---indicating the target laser wavelength and repetition rate could not be mutually satisfied; instead, the algorithm found a compromise: e.g. for the 15~kHz at 1590~nm target, the optimum output that could be found was an 11~kHz pulse train at 1585~nm.
This limitation is related to the finite laser gain profile and intrinsic coupling of the spectral and temporal tuning through the birefringent filtering.

Greater degrees of freedom may enable broader simultaneous tuning of pulse properties and wavelength, beyond these current limits. 
Inclusion of a second polarization controller, for example, after the output coupler would allow control over the polarization state of light launched into the non-PM fiber~\cite{Andral2015}.
Additionally, $>$75~nm of spectral tunability was demonstrated by introducing a variable attenuator into a laser cavity allowing adjustment of the threshold and thus the population inversion, modifying the spectral gain shape~\cite{Meng2015}---this could also be included as a GA control variable to extend the tuning range.
The automated tunability in wavelength and repetition rate suggest that applications such as laser spectroscopy and photoacoustic imaging could be explored to exploit such a Q-switched fiber laser~\cite{Piao2016}.

Finally, we critically discuss the potential of intelligently automated birefringent filtering for the development of self-tuning fiber lasers.
For optimization of a single characteristic, the proposed solution performs well: $>$55~nm tunability was demonstrated, without user intervention or active monitoring of the environment (e.g. temperature compensation).
This could enable reliable automated wavelength tuning of lasers, including pulsed sources (mode-locked or Q-switched) using real SAs.
For multi-characteristic tuning, however, the intrinsic coupling of temporal and spectral properties and finite spectral gain present limitations to the achievable tuning ranges: when demanding a 15~kHz repetition rate, our tuning range was limited to $\sim$20~nm.
Introducing additional degrees of freedom and electronically controlled components is thus an interesting topic for future work.
Compared to alternative approaches, such as in-line active modulators, automated passive filtering to achieve tuning of laser output characteristics represents a novel and potentially simpler/economical route forwards.
We note that miniaturized low-cost optical diagnostics are an area of active research and development; thus, the inclusion of real-time monitoring systems in laser devices will enable further progress in this area.

In conclusion, we have demonstrated the first self-tuning self-Q-switching fiber laser using a genetic algorithm to control intracavity birefringent filtering. 
Self-tuning is possible in the presence of an unregulated environment which has to date prohibited the practical application of artificial saturable absorbers without complex active thermal and mechanical stabilization.
Extending the autonomous tuning range of on-demand laser properties is a future challenge; yet, we believe there is great potential for artificial intelligence in the control of laser systems, not least by harnessing linear and nonlinear polarization effects. 

\subsection*{Acknowledgements}
We thank J. R. Taylor, R. T. Murray and T. H. Runcorn for fruitful discussions and M Squared Lasers for loan of an IceBloc DDTC. We acknowledge support through EPSRC (RIW) \& Royal Academy of Engineering (EJRK) Fellowships.

All data underlying the results presented in this Letter are freely available upon request from the authors.

\end{document}